\documentclass[authoryear,12pt]{elsarticle}

\usepackage{multirow}
\usepackage{epsfig, setspace, pmgraph}%doublespace}
\usepackage{subfigure}
\usepackage{graphicx}
\usepackage{afterpage}
\usepackage{amsmath, amsfonts, amssymb}
\usepackage{dsfont,courier}
\usepackage{verbatim}
\usepackage{longtable}
\usepackage{hyperref}
\usepackage{lineno}

\journal{Journal of Materials Processing Technology}

\begin{document}

\begin{frontmatter}



\title{Analytical Solution of the Tooling/Workpiece Contact Interface Shape During a Flow Forming Operation}


\author[ubc]{M. J. Roy\corref{cor1}}
\ead{majroy@interchange.ubc.ca}
\author[ubc]{D. M. Maijer}
\ead{daan.maijer@ubc.ca}
\author[uwomm]{Robert J. Klassen}
\ead{rklassen@eng.uwo.ca}
\author[uwomm]{J. T. Wood}
\ead{jwood@eng.uwo.ca}
\author[uwocs]{E. Schost}
\ead{eschost@uwo.ca}

\address[ubc]{Dept. of Materials Engineering, The University of British Columbia, Vancouver, BC, Canada V6T 1Z4}
\address[uwomm]{Dept. of Mechanical and Materials Engineering, The University of Western Ontario, London, ON, Canada N6A 5B9}
\address[uwocs]{Computer Science Dept., The University of Western Ontario, London, ON, Canada N6A 5B9}

\cortext[cor1]{Corresponding author, Tel. +1 604 827 5346;
Fax +1 604 822 3619}

\begin{abstract}
Flow forming involves complicated tooling/workpiece interactions. Purely analytical
models of the tool contact area are difficult to formulate, resulting in
numerical approaches that are case-specific.  Provided are the details of an
analytical model that describes the steady-state tooling/workpiece contact area allowing for easy modification of the dominant geometric variables. The assumptions made in formulating this analytical model are validated with experimental results attained from physical modelling.  The analysis procedure can be extended to other
rotary forming operations such as metal spinning, shear forming,
thread rolling and crankshaft fillet rolling.
\end{abstract}


\end{frontmatter}
%\begin{linenumbers}

\section*{Nomenclature}
\begin{description}
    \item[$\alpha$] - the entry angle from workpiece to the roller, also known as the attack angle [$^\circ$] (Fig. ~4)
    \item[$A_{xyz}$] - overall contact area [mm$^2$] (Section \ref{sec:application})
    \item[$A_{xy}$] - $xy$ planar projection of the contact area [mm$^2$] (Section \ref{sec:application})
    \item[$A_{xz}$] - $xz$ planar projection of the contact area [mm$^2$] (Section \ref{sec:application})
    \item[$A_{yz}$] - $yz$ planar projection of the contact area [mm$^2$] (Section \ref{sec:application})
    \item[$\beta$] - the trailing angle from the workpiece to the roller, also known as exit angle or planishing angle [$^\circ$] (Fig. ~4)
    \item[$\delta_D$] - distance between analytical surface and experimental surface nearest neighbour points (Section \ref{sec:validation})
    \item[$\delta_{Di}$] - distance between analytical surface and experimental surface interpolant (Section \ref{sec:validation})
    \item[$d$] - the distance between the center of the mandrel and the center of the roller [mm](Fig. ~4,   $d=R_m+t_f+R+R_r$)
    \item[$f_z$] - axial feed rate of the roller down the face of the cylinder, along the $z$
    direction [mm/min]
    \item[MSE] - Mean Square Error (Eq. \ref{eq:MSE})
    \item[$n$] - the mandrel rate of rotation [revolutions/min]
    \item[$P$] - the roller path pitch or distance traveled axially by the roller in one revolution [mm] ($P=f_z / n$)
    \item[$R$] - roller nose radius [mm] (Fig. ~4)
    \item[$R^*$] - numeric resolution of the solution (Section~\ref{sec:surdef})
    \item[$R_i$] - initial workpiece radius ($R_i = R_m + t_0$)
    \item[$R_m$] - the mandrel radius [mm] (Fig. ~4)
    \item[$R_r$] - the roller radius excluding the radius of the nose [mm] (Fig. ~4)
    \item[$S$] - Intermediate set of radial quantities used to find $X_p$ and $Y_p$ (Eq. \ref{eq:S})
    \item[$\theta_f$] - the angle of contact between the roller and workpiece [rad](Fig. ~2, Eq. \ref{eq:Tf})
    \item[$\theta_i$] - intermediate value of $\theta$ used for the iterative solution of the contact area [rad] (Eq. \ref{eq:Tf})
    \item[$\theta _{\max}$] - maximum angular limit of the solution (Eq. \ref{eq:thetamax})
    \item[$\theta _{1\times R^*}$] - angular coordinates used to define boundary surfaces (Eq. \ref{eq:tcoords})
    \item[$t_0$] - starting material thickness [mm] (Fig. ~4)
    \item[$t_f$] - the final material thickness [mm] (Fig. ~4)
    \item[$X_i$] - $x$ coordinates lying on the instantaneous roller position (Eq. \ref{eq:X_i})
    \item[$x_i$] - $x$ coordinate used for intersection conditioning (Eq. \ref{eq:xi})
    \item[$x_l$] - $x$ coordinate used for intersection conditioning (Eq. \ref{eq:xlower})
    \item[$X_m$] - $x$ coordinate lying on the cylinder defined by $R_i$ (Eq. \ref{eq:X_m})
    \item[$x_{\max}$] - maximum limit in the $x$ direction of the solution (Eq. \ref{eq:xmax})
    \item[$X_p$] - $x$ coordinates lying on the previous roller path (Eq. \ref{eq:Xp})
    \item[$X_s$] - $x$ coordinates within the roller/workpiece contact area (Eq. \ref{eq:Xs})
    \item[$x_u$] - $x$ coordinate used for intersection conditioning (Eq. \ref{eq:xupper})
    \item[$Y_i$] - $y$ coordinates lying on the instantaneous roller position (Eq. \ref{eq:Yi})
    \item[$Y_m$] - $y$ coordinates lying on the cylinder defined by $R_i$ (Eq. \ref{eq:Y_m})
    \item[$y_{\max}$] - maximum limit in the $y$ direction of the solution (Eq. \ref{eq:ymax})
    \item[$Y_p$] - $y$ coordinates lying on the previous roller path (Eq. \ref{eq:Yp})
    \item[$Y_{1\times R^*}$] - $y$ coordinates used to define boundary surfaces (Eq. \ref{eq:ycoords})
    \item[$Y_s$] - $y$ coordinates within the roller/workpiece contact area (Eq. \ref{eq:Ys})
    \item[$z_{1-2}$] - axial limits of the workpiece/roller contact area, $z$ coordinate of the endpoint of contour 1 and starting point of contour 2 (Eq. \ref{eq:lowerlimflat1} and \ref{eq:lowerlimitflat2})
    \item[$z_{1-3}$] - axial limits of the workpiece/roller contact area, $z$ coordinate of the endpoint of contour 1 and starting point of contour 3 (Eq. \ref{eq:zA} to \ref{eq:zD})
    \item[$Z_i$] - $z$ coordinates lying on the instantaneous roller position (Eq. \ref{eq:Zi})
    \item[$z_i$] - $z$ coordinate used for intersection conditioning (Eq. \ref{eq:zi})
    \item[$z_l$] - $z$ coordinate used for intersection conditioning (Eq. \ref{eq:zlower})
    \item[$Z_m$] - $z$ coordinates lying on the cylinder defined by $R_i$ (Eq. \ref{eq:Zm})
    \item[$Z_p$] - $z$ coordinates lying on the previous roller path (Eq. \ref{eq:Zp})
    \item[$Z_{1\times R^*}$] - $z$ coordinates used to define boundary surfaces (Eq. \ref{eq:zcoords})
    \item[$Z_s$] - $z$ coordinates within the roller/workpiece contact area (Eq. \ref{eq:Zs})
    \item[$z_u$] - $z$ coordinate used for intersection conditioning (Eq. \ref{eq:zupper})
\end{description}

%\subsection{Iteration Variables}
%\begin{description}
%\item[$\theta _{i}$] - intermediate tangential limit of contact (Section
% ~\ref{sec:implicit})
%\item[$n$] - number of iterations needed (Section
% ~\ref{sec:implicit})
%\end{description}

\section{Introduction}\label{sec:areaintro}
To determine the energy required to form a component, the size and orientation of the tooling interface on the workpiece is necessary.  While purely analytical models describing this
contact are preferable, they are usually difficult to
attain for complex metal forming processes.  In this study, an analytical approach is presented to model the
tooling/workpiece contact area in an application of rotary forming.
While the present work focuses on an implementation for flow
forming, the applied technique can be applied to other variants of
rotary forming operations such as metal spinning, shear forming,
thread rolling and crankshaft fillet rolling.

Flow forming, a variant of metal spinning, is a process used to
fabricate rotationally symmetrical parts from ductile materials, after
\cite{wong.03}. During flow forming, the workpiece is clamped to a
rotating mandrel and pressed into contact with the mandrel by
rollers. The rollers induce high levels of plasticity in the
workpiece causing it to undergo both reduction in thickness and
axial lengthening. Since the rollers press on only a very small area
of the overall workpiece at any given time, the deformation is
highly localized between the roller and workpiece.  To properly
understand the distribution of this intense local plastic deformation it
is essential to be able to calculate the roller/workpiece
contact area from the geometric parameters that govern the flow
forming process.  In addition, the roller/workpiece contact area is
critical to coupling other experimental findings, such as power
consumption, frictional effects, force, stress and strain
distributions through the workpiece back to geometric process parameters.

In flow forming, the combined mandrel rotation and linear movement of the rollers
induce contact on the workpiece along a helical path.
This helical tool path, coupled with the curved profile of the
rollers leads to a very complicated roller/workpiece contact area.

In terms of related tool contact studies, an important analytical derivation of the workpiece contact in shear spinning was completed by \cite{Chen.05}. However, in a comprehensive review of metal spinning processes, \cite{Music.10} highlighted that the mechanics of flow forming are quite different than shear spinning. This is also true for the contact area formulation as there is little roller penetration into the workpiece and deformation proceeds according to the sine rule.

In terms of flow forming specific research, investigations made by Gur and Tirosh (1982), Singhal et al. (1995), Ma (1993) and Jahazi and Ebrahimi (2000) have proposed analytical models of this contact. Gur and Tirosh (1982) developed the formulation of a planar contact area in each of the primary rolling and extrusion deformation directions in backwards flow forming. Singhal et al. (1995) derived the contact area imposed by tooling in the flow forming of small diameter tubes where the assumption made was that material is assumed to be perfectly plastic, and the tools were assumed rigid. Ma (1993) extended the work of Gur and Tirosh (1982) to derive a critical angle of attack and Jahazi and Ebrahimi (2000) extended the contact formulation made by Gur and Tirosh (1982) to investigate the mechanics in a specific application of flow forming. More recently, Kemin et al. (1996), Xu et al. (2001) and Hua et al. (2005) have developed Finite Element (FE) models of single roller flow forming. In each of these studies, contact was modeled explicitly within each respective FE model. Furthermore, with the exception of the work by \cite{xu.01} and \cite{Hua.05}, all
previous works have made assumptions concerning the roller/workpiece
contact geometry that do not necessarily reflect the actual contact
during flow forming. These assumptions include:

\begin{enumerate}
\item Idealized roller geometry (i.e. no blending
radii) (Fig. ~1);
\item The use of two-dimensional treatments that do not account for the three-dimensional aspects of the workpiece
contact;
\item Not considering the influence of prior forming steps (i.e. roller path overlap) on the instantaneous roller/workpiece contact area.
\end{enumerate}

The most successful technique for modelling the roller/workpiece contact area, and
other facets of the flow forming process, has been through FE
analyses.  \cite{xu.01} addressed items 1 and 2
listed above in their work to numerically calculate the
roller/workpiece of a single roller flow forming. However, \cite{xu.01} did
not give the details of their calculation of the contact area, nor
did they specifically address item 3. \cite{Hua.05} has developed a thorough 3-D FE model that addresses all three items, but an FE approach is still limited to case-by-case application involving extensive pre-processing and explicit geometric modelling. An analytical solution provides a solution with significantly lower effort.
In the present work, a generalized solution is developed for the
roller/workpiece contact area during a single roller flow forming
operation that accommodates items 1 to 3.  To accomplish this, the
following assumptions are made:
\begin{enumerate}
\item The single roller flow forming process proceeds under steady state conditions. The final and starting thickness, mandrel rotation and feed rate are constant.
\item The deformation response of the workpiece is perfectly plastic.  Elastic effects are not considered.
\item Volume of the flow formed workpiece is conserved outside the tool interface.
\item No material build-up occurs in front of the roller as the workpiece conforms completely with the rigid roller.
\end{enumerate}

\section{Contact Solution}\label{section:Description}
During flow forming, the roller contacts the workpiece along a path
having a constant pitch (Fig. 1). The profile of the roller can be
divided into three regions; the entry region, the nose region and
the exit region.  These regions dictate the size and shape of the
roller/workpiece contact area.  The contact area is
bounded by three contours: the tangential exit contour, the axial
entry contour and the axial exit contour, labeled 1-3 respectively
in Fig. 2(a).  The contact area extends angularly from the tangential
exit contour ($\theta=0$) through to $\theta = \theta _f$ (Fig. 2(b)).
\begin{figure}
\centering
\label{fig:1}
      \includegraphics[width=4 in]{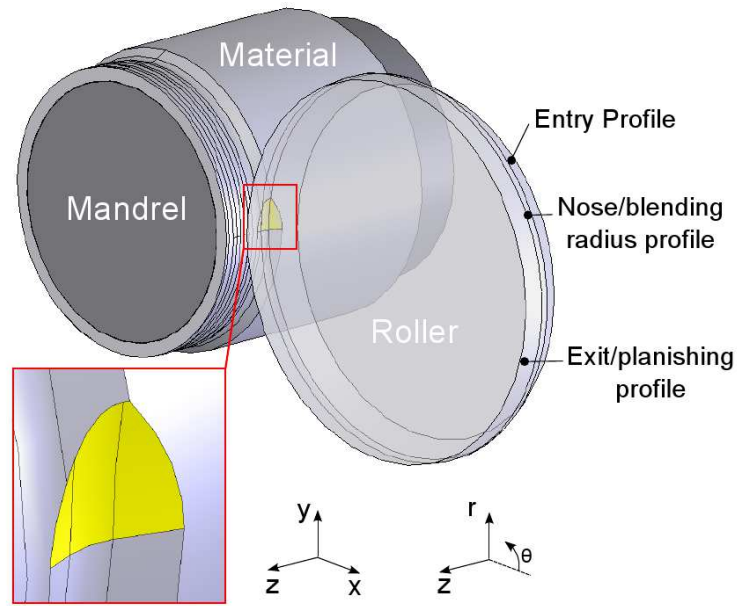}
      \caption[]{Single roller contact in flow forming showing the mandrel and key roller profiles.}
\end{figure}
\begin{figure}%[htp]
     \centering
     \subfigure[]{\label{fig:2a}
          \includegraphics[width=2.35 in]{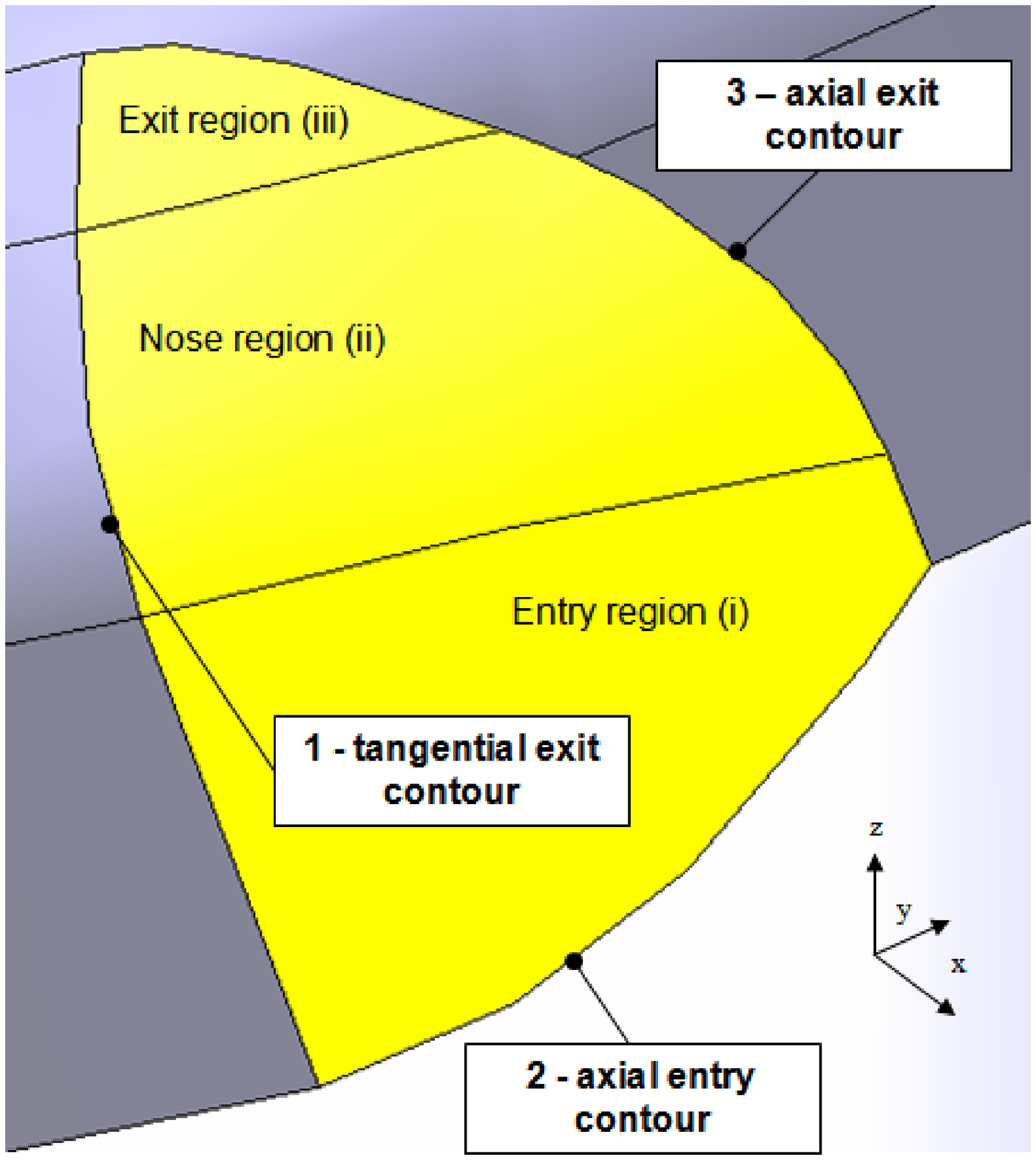}}
     \hspace{.3in}
     \subfigure[]{\label{fig:2b}
          \includegraphics[width=2.35 in]{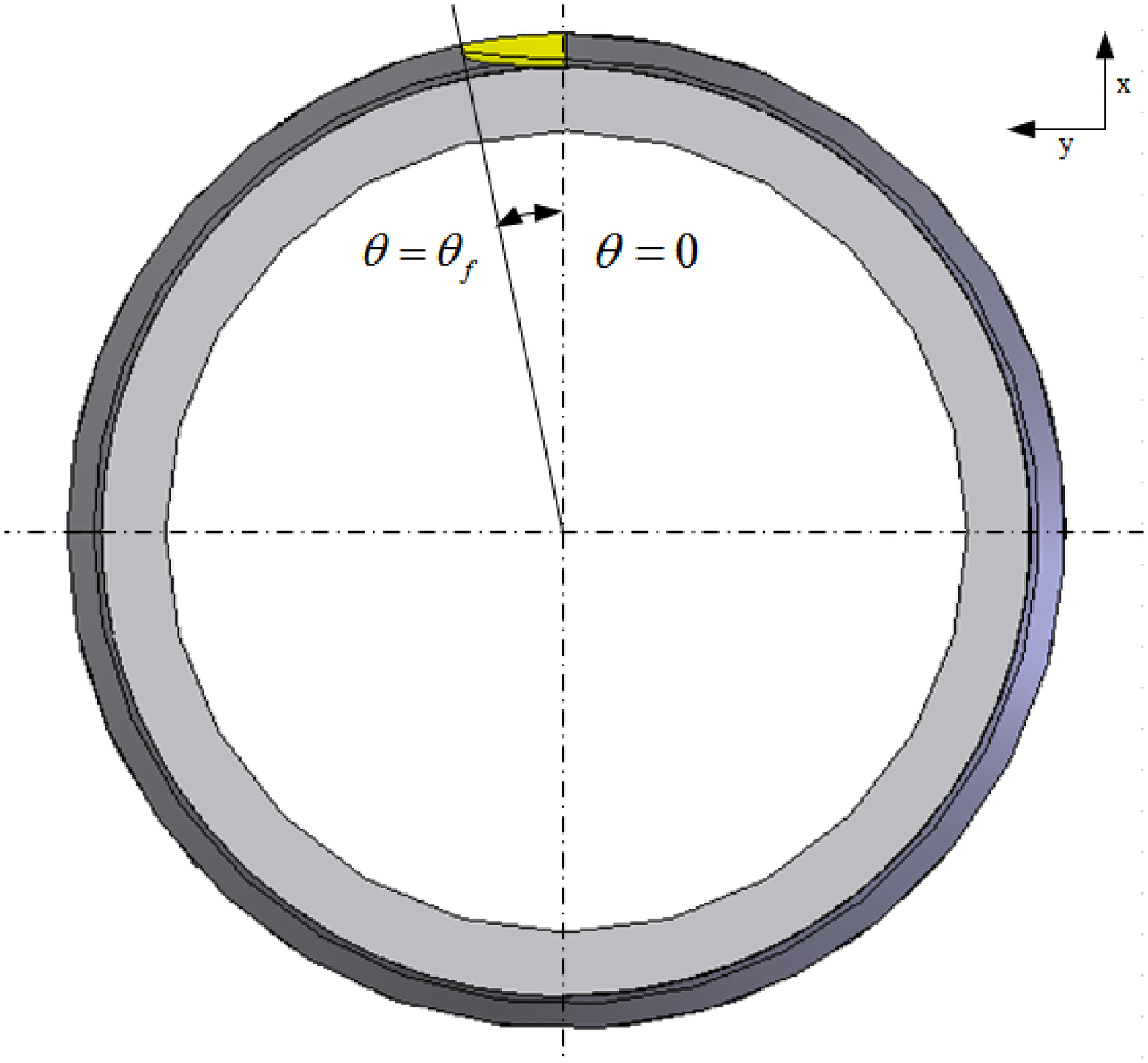}}
\caption[]{(a) Detail of roller contact showing different zones and contour numbers.  (b) Contact extends angularly
  from 0 to $\theta_f$.}
     \label{fig:2}
\end{figure}
If the roller has an archetypal flow forming profile similar to that shown in Fig. 1 with
distinct flat entry and exit regions and a blending radius
between the two that creates a nosed roller, the final contact area
is dependent on six surfaces (Fig. 3). Contour 1, and the starting
points of contours 2 and 3 (Fig. 2) can be calculated directly as
they lie exclusively on the $xz$-plane. The \emph{a priori}
$z$-coordinates of the extents of contour 1 define the axial limits
of the roller/workpiece contact area.  Once the a solution has been found for the starting and ending points of contour 1 (by definition the starting points of contours 2 and 3),
the common end point of contours 2 and 3 is then solved
using an iterative technique.
\begin{figure}
\centering
\label{fig:3}
      \includegraphics[width=4 in]{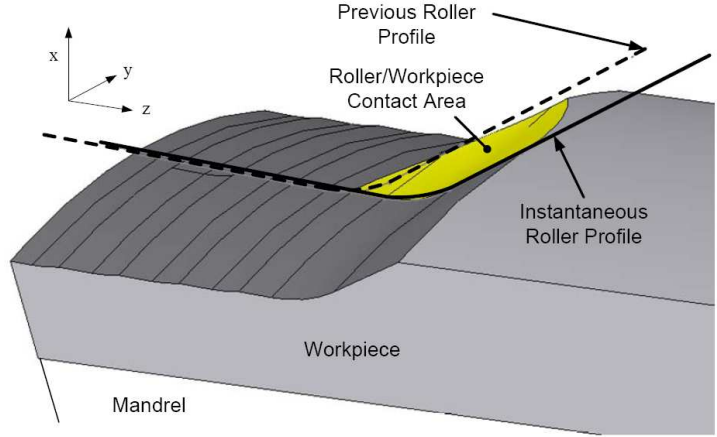}
      \caption[]{Roller profiles deciding the instantaneous contact area during flow
forming.  There are six in total: the nosed region of the roller
from the previous workpiece rotation, the entry region of the roller
from the previous workpiece rotation, the instantaneous roller exit
region, the instantaneous roller nosed region, the instantaneous
roller entry region and the outer surface of the unformed workpiece.}
\end{figure}
\subsection{Axial Limits}\label{sec:axiallimits}
It is first necessary to calculate the axial limits of contact by determining the endpoints of contour 1.
Contour 2 is a function of the
instantaneous roller contact with the workpiece at pitch $P=0$.
Contour 3 is a function of the
instantaneous roller contact on the material as well as the tool
contact on the workpiece one revolution of the mandrel beforehand, at
$P=f_z/n$. Contour 1 exists solely on
the $xz$ plane and is bound by the points of intersection with
contours 2 and 3.  Contour 1 is both
dependent on roller geometry and the roller path pitch, $P$.
There are four possible conditions describing the intersection of
the current roller position with that of its position on the
previous mandrel revolution (Fig. 4).
\begin{figure}
\centering
\label{fig:4}
      \includegraphics[width=4 in]{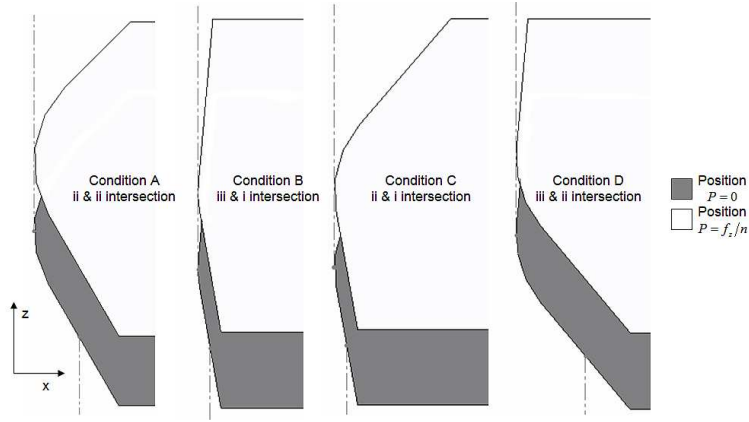}
      \caption[]{Upper limit conditions of the contact area. Position $P=0$ corresponds to the instantaneous roller position and
  position $P=f_z/n$ corresponds to the roller at one mandrel revolution beforehand.  The upper
endpoint of contour 1 can occur within the nosed region of the
roller on both the instantaneous position and the position on the
previous mandrel revolution (condition A). It can also occur at the
intersection of the exit/entry profiles (condition B), the
nosed/entry profiles (condition C) or the exit/nosed profiles
(condition D) of the instantaneous and the previous roller
positions.}
\end{figure}
Calculation of the location of
the upper end point of contour 1 for the four conditions shown in
Fig. 4 is accomplished through comparison of the endpoints of the roller
nose profile on the $xz$ plane.  For comparison purposes, the local
coordinate system is moved on the $x$ axis from the global origin by
$d-R_r-R$ (Fig. 5). The $x$ and $z$ coordinates of the upper end point
of the nosed region of contour 1, $x_u$ and $z_u$ (Fig. 5):

\begin{figure}
\centering
\label{fig:5}
      \includegraphics[width=4 in]{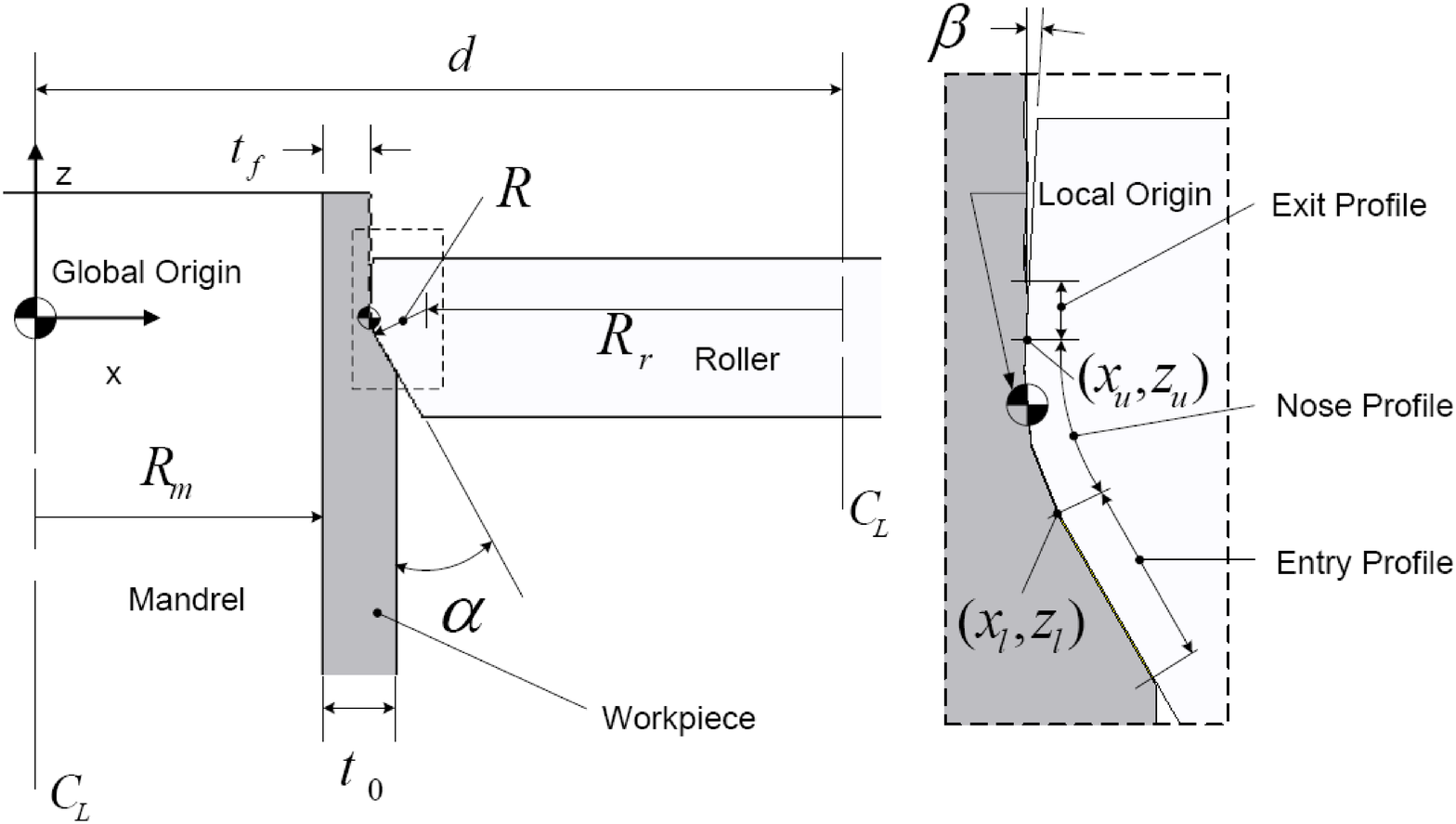}
      \caption[]{Two dimension cutaway of a single flow forming operation showing critical
  geometric variables and forming zones.}
\end{figure}

\begin{equation}\label{eq:xupper}
 x_u  = R\left( {1 - \cos \beta } \right)
\end{equation}

\begin{equation}\label{eq:zupper}
z_u  = R\sin \beta
\end{equation}

\noindent For the lower $x$ and $z$ coordinates of the end point of the
nose region of contour 1, $x_l$ and $z_l$ (Fig. 5):

\begin{equation}\label{eq:xlower}
 x_l  = R\left( {1 - \cos \alpha } \right)
\end{equation}

\begin{equation}\label{eq:zlower}
 z_l  =  - R\sin \alpha
\end{equation}

\noindent The entry profile of the previous roller path and the
instantaneous exit profile of the roller will occur at $x_i$ and
$z_i$.  These are expressed as:

\begin{equation}\label{eq:xi}
\begin{split}
x_i = \frac{{R(\sin \alpha \cos \beta + \sin \beta \cos
\alpha - \sin \beta  - \sin \alpha ) + P\sin \alpha \sin \beta
}}{{\cos \beta \sin \alpha  + \sin \beta \cos \alpha }}
\end{split}
\end{equation}

\begin{equation}\label{eq:zi}
\begin{split}
z_i = \frac{{R(\cos \alpha  - \cos \beta ) + P\cos \beta \sin \alpha
}}{{\cos \beta \sin \alpha  + \sin \beta \cos \alpha }}
\end{split}
\end{equation}

\noindent The values of $x_u$, $x_l$, $x_i$, $z_u$, $z_l$ and $z_i$
can be compared to identify which contact condition shown in Fig. 4
applies.  The conditions and the relationships that must be
simultaneously satisfied are shown in Table \ref{table:conditions}.

\clearpage
\begin{table}[h!]
  \centering
    \caption{Upper axial limits of contact}\label{table:conditions}
    \begin{singlespacing}
    \begin{tabular}{ccc}
    \hline
    \multirow{2}{*}{\textbf{Condition}}&\multirow{2}{*}{\textbf{Relationships}}&\textbf{Intersection}\\
    & & ($P=0$/$P=f_z/n$)\\
    \hline
    \multirow{4}{*}{A}&$z_l+P \leq z_i$&\multirow{4}{*}{Nosed region/Nosed region}\\
    &$z_u>z_i$& \\
    &$x_l>x_i$& \\
    &$x_u>x_i$& \\
    \hline
    \multirow{4}{*}{B}&$z_l+P \geq z_i$&\multirow{4}{*}{Exit region/Entry region}\\
    &$z_u<z_i$& \\
    &$x_l<x_i$& \\
    &$x_u<x_i$& \\
    \hline
    \multirow{4}{*}{C}&$z_l+P \geq z_i$&\multirow{4}{*}{Nosed region/Entry region}\\
    &$z_u>z_i$& \\
    &$x_l<x_i$& \\
    &$x_u>x_i$& \\
    \hline
    \multirow{4}{*}{D}&$z_l+P \leq z_i$&\multirow{4}{*}{Exit region/Nosed region}\\
    &$z_u<z_i$& \\
    &$x_l>x_i$& \\
    &$x_u<x_i$& \\
    \hline
\end{tabular}
\end{singlespacing}
\end{table}

Once the proper contact condition is determined, a solution for the $z$ coordinate for the
upper end point of contour 1, as well as the upper axial limit of
the solution space, $z_{1-3}$, is possible. Solving $z_{1-3}$ for
the appropriate condition A through D:

\begin{equation}\label{eq:zA}
z_{1 - 3} \left\{ {\rm A} \right\} = \frac{P}{2}
\end{equation}

\begin{equation}\label{eq:zB}
z_{1 - 3} \left\{ {\rm B} \right\} = z_i
\end{equation}

\begin{equation}\label{eq:zC}
z_{1 - 3} \left\{ {\rm C} \right\} = \left\{
\begin{array}{l}
 z = \sqrt {R^2  - (x - R)^2 }  \\
 z = \frac{x}{{\tan \alpha }} + \left( {\frac{{x_l }}{{\tan \alpha }} + z_l  + P} \right) \\
 \end{array} \right\}
\end{equation}

\begin{equation}\label{eq:zD}
z_{1 - 3} \left\{ {\rm D} \right\} = \left\{
\begin{array}{l}
 z = P - \sqrt {R^2  - (x - R)^2 }  \\
 z = \frac{x}{{\tan \beta }} + z_u  - \frac{{x_u }}{{\tan \beta }} \\
 \end{array} \right\}
\end{equation}

\noindent In conditions C and D, $z_{1-3}$ is expressed as the solution that
satisfies the two equations for $z$.

The lower end point of contour 1, $z_{1-2}$, occurs at the intersection of the profile of the
instantaneous roller position and the cylinder with radius $R_i$.  This intersection depends on roller geometry and
the depth that it penetrates into the workpiece. Either the roller
intersects the workpiece at the flat entry region (condition I) or the
nosed region (condition II). If the roller intersects at the flat entry
region:

\begin{equation}\label{eq:xlowerlimittest}
x_l  \leq t_f
\end{equation}

\noindent where $x_l$ is given by Eq. \ref{eq:xlower}. Otherwise,
condition II prevails and the roller intersects within the nosed
region.  The solution for this intersection point yields the following
expressions for the lower end point of contour 1, and the lower
axial limit of the solution space.  This value, $z_{1-2}$, for each
condition is given as:

\begin{equation}\label{eq:lowerlimflat1}
z_{1 - 2} \{ {\rm I}\}  =  - \left( {\frac{{t_0  - t_f  + R(\sec
\alpha  - 1)}}{{\tan \alpha }}} \right)
\end{equation}

\begin{equation}\label{eq:lowerlimitflat2}
z_{1 - 2} \{ {\rm II}\}  =  - \sqrt {\left( { - t_0 ^2  + 2t_0 t_f -
t_f ^2  + 2Rt_0  - 2Rt_f } \right)}
\end{equation}

\subsection{Solution Boundaries}
Thus far, the region where the tool contact resides has been
explicitly bound in the axial direction between $z_{1-3}$ and
$z_{1-2}$. The following describes how the components needed for a
computation of the full three dimensional contact area are developed. These components are
the maximum angular limit that the solution space can be defined by
and the surfaces that bind the solution space of the instantaneous contact.

\subsubsection{Maximum Angular Limit}\label{sec:maxradlim}
Contours corresponding to the ones described in Section
\ref{section:Description} that pass through the axial limits,
$z_{1-3}$ and $z_{1-2}$, are formulated to extend angularly from
$\theta=0$ to $\theta=\theta_{\max}$.  This value is the absolute
maximum value that $\theta_f$ can be, corresponding to $P\simeq 0$.
The extremal point at angle $\theta_{\max}$ lies on the $xy$ plane
at $z=0$; its coordinates are obtained using the same derivation as
the general solution for the contact of two circles using the global datum. The first circle is one
centered at $x,y=0$ with a radius of $R_i$ and the other is at a distance
$x=d$ and $y=0$ with a radius of $R_r+R$.

\begin{equation}\label{eq:thetamax}
\theta _{\max }  = \arctan \left( {\frac{{y_{\max } }}{{x_{\max }
}}} \right)
\end{equation}

\noindent with $x_{\max}$ and $y_{\max}$ as follows:

\begin{equation}\label{eq:xmax}
x_{\max }  = \frac{{d^2  - (R_r+R) ^2  + R_i ^2 }}{{2d}}
\end{equation}

\begin{equation}\label{eq:ymax}
y_{\max }  =  \frac{{\sqrt{4d^2 R_i ^2  - \left( {d^2  - (R_r+R) ^2
+ R_i ^2 } \right)^2 } }}{2d}
\end{equation}

\subsubsection{Surface Definitions}\label{sec:surdef}
As described in Section \ref{section:Description}, there are six different surfaces that define the area.  Due to the system
complexity, a numerical technique is employed to generate these
boundary surfaces in three dimensions. In this technique, the
overall solution space is represented by a finite number of points
or nodes. For example, if the solution space was broken up into
$20 \times 20 \times 20$ uniformly spaced points, then the resolution,
$R^*$, of the space would be 20.  Figure 6 shows the effect on the solution by increasing or decreasing $R^*$.
\begin{figure}%[htp]
     \centering
     \subfigure[]{\label{fig:6a}
          \includegraphics[width=2.35 in]{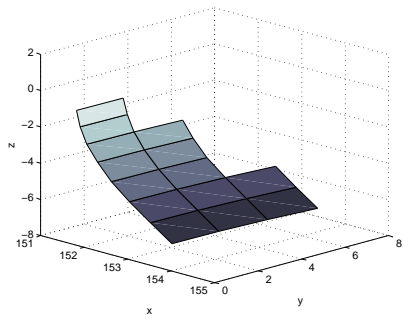}}
     \hspace{.3in}
     \subfigure[]{\label{fig:6b}
          \includegraphics[width=2.35 in]{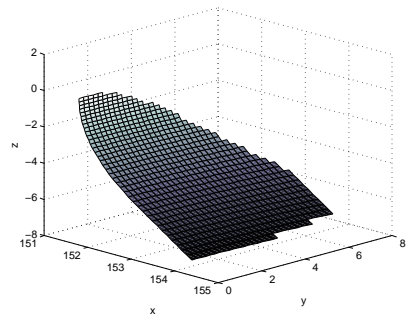}}\\
          %\vspace{.3}
          \subfigure[]{\label{fig:6c}
          \includegraphics[width=2.35 in]{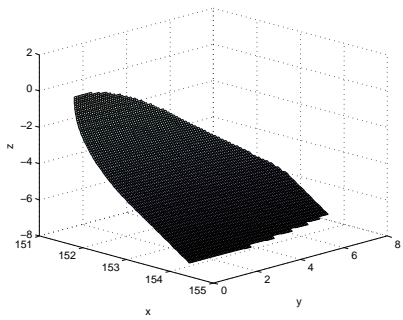}}
\caption[]{Graphical progression of the iterative algorithm used to solve the contact
area. Starting with a coarse result shown in (a) with $R^*=10$, with and resolution increased
in (b) to $R^*=40$, and finally with a high resolution answer in
(c) with $R^*=160$.}
     \label{fig:6}
\end{figure}
The following are the definitions of the
arrays of coordinates used to define the boundary surfaces as
functions of $R^*$, where initially $n=R^*$ and decays for every term included
in the array until $n=1$.

\begin{equation}\label{eq:zcoords}
Z_{1{\rm x}R^* }(n) =   z_u  - \left( {R^*  - n} \right)\left(
{\frac{{z_u  - z_l }}{{R^*  - 1}}} \right) \quad \text{ for } n=R^*,R^*-1,\ldots 1
%\begin{split}
%Z_{1{\rm x}R^* }(n) =   \quad &z_u  - \left( {R^*  - n} \right)\left(
%{\frac{{z_u  - z_l }}{{R^*  - 1}}} \right),\\
%& z_u  - \left( {R^*  - n- 1} \right)\left( {\frac{{z_u - z_l
%}}{{R^* - 1}}} \right),\\
%& z_u  - \left( {R^*  - n - 2} \right)\left( {\frac{{z_u  - z_l
%}}{{R^*  - 1}}} \right), \ldots z_u - \left( {R^* - 1} \right)\left(
%{\frac{{z_u  - z_l }}{{R^*  - 1}}} \right)
%\end{split}
\end{equation}

\begin{equation}\label{eq:ycoords}
Y_{1{\rm x}R^* }(n)= \left( {R^*  - n} \right){\frac{{R_i\sin \theta _f }}{{R^*  - 1}}} \quad \text{ for } n=R^*,R^*-1,\ldots 1
%\begin{split}
%Y_{1{\rm x}R^* }  =    \quad &0, \\
%& \left( {R^*  - n - 1} \right){\frac{{R_i\sin \theta _f }}{{R^*  - 1}}}, \\
%& {\left( {R^*  - n - 2} \right)\left( {\frac{{R_i\sin \theta _f
%}}{{R^* - 1}}} \right)},\ldots {\left( {R^*  -1} \right)\left(
%{\frac{{R_i\sin \theta _f }}{{R^* - 1}}} \right)}
%\end{split}
\end{equation}

\begin{equation}\label{eq:tcoords}
\theta _{1{\rm x}R^* }(n) = \left( {R^*  - n} \right){\frac{{\theta _f }}{{R^*  - 1}}}\quad \text{ for } n=R^*,R^*-1,\ldots 1
%\begin{split}
%\theta _{1{\rm x}R^* }  = \quad &0,\\
%& \left( {R^*  - n - 1} \right){\frac{{\theta _f }}{{R^*  - 1}}},\\
%& {\left( {R^*  - n - 2} \right)\left( {\frac{{\theta _f }}{{R^*  - 1}}} \right)}, \ldots  {\left( {R^*  - 1} \right)\left( {\frac{{\theta _f }}{{R^*  - 1}}} \right)}  \\
%\end{split}
\end{equation}

Now that the arrays of points are formulated, the boundary surfaces
can be formed as $m=R^*$ by $n=R^*$ square matrices for each
direction through space to form $Z_{_{m,n}}$, $Y_{_{m,n}}$ and $\theta_{_{m,n}}$.  The $m$ direction of these matrices
is a solution for $z$ at a given value in the $n$ direction of $y$
or $\theta$, corresponding to the coordinate arrays given in
Equations \ref{eq:zcoords}, \ref{eq:ycoords} and \ref{eq:tcoords}. The following are functions of discrete entries in the matrices that define the boundary surfaces.

Starting with the instantaneous roller position:
\begin{equation}\label{eq:X_i}
X_{i_{m,n}} = \\
\left\{
\begin{array}{lr}
d + \frac{{R(\cos \beta  - \cos \alpha )}}{{\cos \alpha \cos \beta
}} - (\phi  - Z_{_{m,n}}\tan \beta )\sqrt {1 - \frac{{R_i^2 \sin ^2
\theta_{_{m,n}} }}{{\phi ^2 }}}
& Z_{_{m,n}} > z_u \\
d - (\phi  - Z_{_{m,n}}\tan \beta )\sqrt {1 - \frac{{R_i^2 \sin ^2
\theta_{_{m,n}} }}{{\phi ^2 }}}
 & Z_{_{m,n}} < z_l\\
d - \sqrt {\left( {\left( {R^2  - Z_{_{m,n}} ^2 } \right) + R_r ^2 }
\right)  -
Y_{_{m,n}} ^2 }
 & z_u \geq Z_{_{m,n}} \geq z_l\\
\end{array}
\right.
\end{equation}
\noindent where $\phi = R_r  + \frac{R}{{\cos \alpha }} $.
$Z_{i_{m,n}} $ and $Y_{i_{m,n}}$ remain the matrices of dimension $R^*$ by $R^*$:
\begin{equation}\label{eq:Yi}
Y_{i_{m,n}}=Y_{_{m,n}}
\end{equation}
\begin{equation}\label{eq:Zi}
Z_{i_{m,n}}=Z_{_{m,n}}
\end{equation}
The cylindrical surface defined by $R_i$ that describes the outer surface of the
workpiece formed around the mandrel:
\begin{equation}\label{eq:X_m}
X_{m_{m,n}}  = R_i \cos \theta_{_{m,n}}
\end{equation}
\begin{equation}\label{eq:Y_m}
Y_{m_{m,n}}  = R_i \sin \theta_{_{m,n}}
\end{equation}
\begin{equation}\label{eq:Zm}
Z_{m_{m,n}}  = Z_{i_{m,n}}
\end{equation}
For the previous roller path, the surfaces are most easily defined in radial expressions, and
translated to cartesian coordinates. This radial quantity, $S$ is
defined as:
\begin{equation}\label{eq:S}
S = \\
\left\{
\begin{array}{lr}
d + \frac{{R(\cos \beta  - \cos \alpha )}}{{\cos \alpha \cos \beta
}} - \ldots\\
\ldots\left( {R\tan \beta \left( {\frac{\phi }{{\tan \beta }} -
Z_{_{m,n}} + P\left( {1 - \frac{\theta_{_{m,n}}}{{2\pi }}} \right)}
\right)} \right)\phi ^{ - 1}
& Z_{_{m,n}} > z_u+P \\
d + \tan \alpha \left( { - \frac{\phi }{{\tan \beta }} - Z_{_{m,n}}  +
P\left( {1 - \frac{{\theta _{m,n} }}{{2\pi }}} \right)} \right)
 & Z_{_{m,n}} < z_l+P\\
d - Rr - \sqrt {R^2  - \left( {Z_{_{m,n}}  - P\left( {1 -
\frac{{\theta_{_{m,n}} }}{{2\pi }}} \right)} \right)^2 }
 & z_u +P \geq Z_{_{m,n}} \geq z_l+P\\
\end{array}
\right.
\end{equation}
\noindent where $\phi = R_r  + \frac{R}{{\cos \alpha }} $.
Converting these radial quantities into cartesian coordinates
results in:
\begin{equation}\label{eq:Xp}
 X_{p_{m,n}}  = S \cos \theta_{_{m,n}}
\end{equation}
\begin{equation}\label{eq:Yp}
  Y_{p_{m,n}}  = S \sin\theta_{_{m,n}}
\end{equation}
\begin{equation}\label{eq:Zp}
Z_{p_{m,n}} = Z_{m_{m,n}}  = Z_{i_{m,n}}
\end{equation}
\subsubsection{Contact Surface Solution}\label{sec:sursol}
Once the boundary surfaces have been defined, the contact surface
can be solved.  The contact surface points reside within the
instantaneous roller position definition, bound by the intersections
with the workpiece surface and the previous roller path surface.  In
order to determine which points lie within these boundaries, the
matrices containing the surfaces are parsed with logical operators
and placed in a new set of matrices corresponding to the $x$, $y$
and $z$ points lying on the contact surface.

For the $x$-coordinates of the contact surface, $X_s$ belongs
to the set of coordinates corresponding to:
\begin{equation}\label{eq:Xs}
X_s  = X_i\{ X_{i_{m,n}}  \leq X_{m_{m,n}} \wedge X_{i_{m,n}}  \leq
X_{p_{m,n}} \}
\end{equation}

\noindent Similarly, for the $y$-coordinates, $Y_s$:

\begin{equation}\label{eq:Ys}
Y_s  = Y_i\{ Y_{i_{m,n}}  \leq Y_{m_{m,n}} \vee Y_{i_{m,n}}  \leq
Y_{p_{m,n}} \}
\end{equation}

\noindent where `$\wedge$' is the conjunction (and) operator, and
`$\vee$' is the disjunction (or) operator. Finally, for the $z$
coordinates, $Z_s$:
\begin{equation}\label{eq:Zs}
Z_s  = Z_i\{ Z_{i_{m,n}}  \leq Z_{p_{m,n}} \}
\end{equation}

Each of these conditions must be \emph{simultaneously} satisfied for
$X_s$, $Y_s$ and $Z_s$.  While all $x$-coordinates will lie between
the cylinder defined by $R_i$ and the previous tool path (leading to
the `$\wedge$' operator), due to the curvature of the
instantaneous roller position versus the counterclockwise curvature
of the workpiece, the $y$-coordinates require the `$\vee$' operator
for the same comparison.

\subsection{Iterative Calculation of the Contact Length}\label{sec:implicit}
The contact area extends from $\theta=0$ on the $xz$ plane to some
final radial distance at $\theta=\theta_f$.  Depending on the geometry involved, the resolution selected and the initial value for $\theta$ selected, it is possible that only half of the coordinates in $X_i$, $Y_i$ and $Z_i$ correspond to
$X_s$, $Y_s$ and $Z_s$. To overcome this, it is necessary to iterate on the initial value of $\theta$ defining the solution space via the following method.  The initial solution without any iteration ($i=0$) has all of the surfaces defined
extending axially through the endpoints $z_{1-3}$ and $z_{1-2}$ of
contour 1 (Fig. 2) and angularly from $\theta=0$ to $\theta_{i=0}$,
corresponding according to
$\theta_{i=0}=\theta_{\max}$. Once the first values for $X_s$, $Y_s$ and $Z_s$ are available, a
preliminary calculation for $\theta_f$ is possible, residing within
$Y_s$ as per Eq. \ref{eq:Tf}.

\begin{equation}\label{eq:Tf}
\theta_f \approx \theta _{i}  = \sin \left( {\frac{{\max \left( {Y_s }
\right)}}{{R_i }}} \right)
\end{equation}

After priming, $\theta_{i=0}$ is probably close to the actual value
of $\theta_f$.  However, the solution space that defines $\theta_{i=1}$ is
spread uniformly across the range of $0<\theta < \theta_{\max}$ and therefore the solution should be retried at a smaller value than $\theta=\theta_{max}$. This can be continued $n$ times until $\theta_{i=n}$ is a large percentage ($\approx$ 95\%) of $\theta_{i=n-1}$ to ensure the most accurate result at a given value of $R^*$.

Once the coordinates for the solution surface have been solved, the
total surface area can be readily calculated.  This can be accomplished
through tessellating or meshing the coordinates within $X_s$, $Y_s$ and $Z_s$, then using Gaussian quadrature or a brute force method to find the area of each element and summing.

\section{Experimental Validation}\label{sec:validation}
Flow forming is typically applied to forming metallic components involving high speeds, feed rates and forming forces. Due to the nature of the process, it is difficult to completely stop the process at a particular point in time in a safe manner such that the instantaneous forming zone is preserved. Therefore, the forming speeds, forces and feed rates must be significantly reduced in order to study the tooling/workpiece interaction in flow forming.

In order to validate the analytical model above, a physical model of flow forming was developed using Plasticine conforming to ASTM D-4236. The suitability of using Plasticine and similar compounds to model metal forming processes has been established by \cite{Sofuoglu.00} as well as \cite{Pertence.97}. All material preparation steps detailed by \cite{Sofuoglu.00} were followed. The contact condition used for validation was designed to provide geometry that other modelling efforts have failed to address. Specifically, the contact of a nosed roller with a nose/nose entry condition of type `AII' (Section \ref{sec:axiallimits}). Unlike the other contact conditions, this type is the most complex in terms of curvature as there is no straight/linear section appearing anywhere on the surface.

%\subsection{Experimental Process Parameters}
The tooling used was a smooth mandrel with $R_m$ = 69.33 mm and a roller with $R_r$ = 56.23 mm, $R$ = 5.00 mm, $\alpha=45^{\circ}$ and $\beta=60^{\circ}$. These were installed on a lathe with a thread-cutting feed set such that $P = 2.54$ mm/rev. The mandrel was dusted with talcum powder and Plasticine of a uniform thickness $t_o=8.05$ mm was set on it. The outer surface of the Plasticine was also dusted with talcum powder. The roller was brought into contact with the Plasticine such that $t_f = 6.17$ with $n = 5$ rev/min. The steel and Plasticine process components were all measured with standard contact measurement apparatus with an accuracy of $\pm 0.01$ mm. This reduction level was selected such that it maximized the size of the contact patch with the given tooling and minimized bulging of material ahead of the roller. Despite these precautions, there was some build-up of material ahead of the roller. The forming was stopped with a brake after two full rotations of the mandrel resulting in the contact patch of the roller mid-forming.

\subsection{Experimental Surface Measurement}
The workpiece and mandrel were removed from the lathe and imaged using a FARO Laser ScanArm\footnote{FARO Technologies, Inc.} controlled with Geomagic Studio 9\footnote{Geomagic Inc.} software. The resulting 3-D point cloud of the experimental flow formed tooling/workpiece surface was used to compare with the analytical solution. In an effort to assess the accuracy of the surface scan, both the mandrel and the contact patch were scanned together. The mandrel portion of the scan gave a mean of $R_m=69.334$ mm over 16,000 points. Defining the difference between the scan and the contact measurement (69.331 $\pm$ 0.013 mm) of the mandrel radius as error, the Mean Square Error (MSE) of the scan was 0.0366 mm. Equation \ref{eq:MSE} is the relationship used for the MSE calculation where $n$ is the number of samples and $\delta_D$ is the difference in distance. This measurement from the scan in comparison to the contact measurement made with a micrometer indicates that the FARO scan accuracy is on the same order as a conventional contact measurement.
\begin{equation}\label{eq:MSE}
\text{MSE}^2=\frac{1}{n}\sum_n \delta_D^2
\end{equation}

\subsection{Comparison to the Analytical Solution}
The analytical surface was generated from process geometry listed above and was used to generate a surface of the contact interface in the form of a 3-D point cloud for comparison.  The two point clouds (analytical and experimental) were then overlaid. Since the origin of the experimental point cloud was unknown, a procedure was developed to determine the correct relative position and register the points of the experimental cloud.  The final location was arrived at by axial and angular perturbations of the analytical surface origin. An assessment of the degree of fit of the analytical surface was found by calculating the nearest neighbour points on the experimental surface. For each point on the analytical surface, the experimental point cloud was parsed until a point with minimum distance was found. The distance between the analytical point and the nearest neighbour experimental point was defined as $\delta_D$. The mean of the distance between analytical and experimental points ($\delta_D$), and the Mean Square Error (MSE) were minimized by translating the analytical surface in the axial and angular directions to determine the final location. Table \ref{table:changeplace} shows a summary of how these parameters change by moving the origin of the analytical surface relative to the experimental one. In order to compare the surfaces, an intermediate surface was linearly interpolated through both the experimental and analytical point clouds and the distance between the surfaces ($\delta_{Di}$) was found.
\begin{table}
\centering
\caption{Mean $\delta_D$ and MSE results of analytical to experimental surface with circumferential and axial perturbation}\label{table:changeplace}
    \begin{singlespacing}
    \begin{tabular}{lcc}
    \hline
    \textbf{Condition} & \textbf{Mean $\delta_D$} & \textbf{MSE}\\
    \hline
    Baseline & 0.1614 & 0.4017\\
    \hline
    +0.4 mm circumferential & 0.1770 & 0.4207\\
    -0.4 mm circumferential & 0.1648 & 0.4059\\
    \hline
    +0.4 mm axial & 0.1630 & 0.4037\\
    -0.4 mm axial & 0.2607 & 0.5106\\
    \hline
    %+0.8 mm circumferential & 0.2085 & 0.4566\\
%    -0.8 mm circumferential & 0.1801 & 0.4243\\
%    \hline
%    +0.8 mm axial & 0.2499 & 0.4999\\
%    -0.8 mm axial & 0.4005 & 0.6328\\
%    \hline
\end{tabular}
\end{singlespacing}
\end{table}
%\subsection{Evaluation}

Figure 7 shows a comparison of the experimental point cloud and the corresponding analytical contact patch. Due to the surface overlap, the analytical surface is depicted as being offset along the $x$ axis by 15 mm from the experimental surface. At the best fit position, the analytical surface is qualitatively indistinguishable from the experimental surface. Figures 8 and 9 show the 2-D plots of the analytical profile of the roller in relation to corresponding nearest neighbour experimental points on the $xy$ plane (Fig. 8) and the $zx$ plane (Fig. 9). These two plots show the proximity of the analytical surface to the experimental one. Outside of the area influence of the roller, there is a lack of coherence with the analytical surface: this is due to the inherent bulging of material ahead of the roller both axially and tangentially.
\begin{figure}
\centering
\label{fig:7}
      \includegraphics[width=4 in]{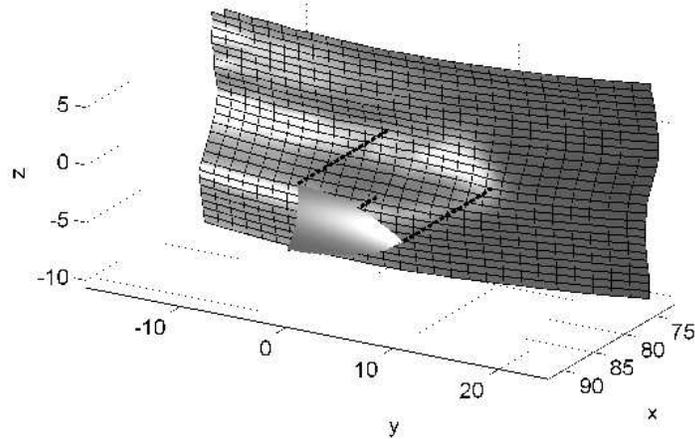}
      \caption[]{Numerical surface generated from the FARO LaserScan point data with the associated analytical surface offset 15 mm along the $x$ axis.}
\end{figure}
\begin{figure}%[htp]
     \centering
     \subfigure[]{\label{fig:8a}
          \includegraphics[width=4 in]{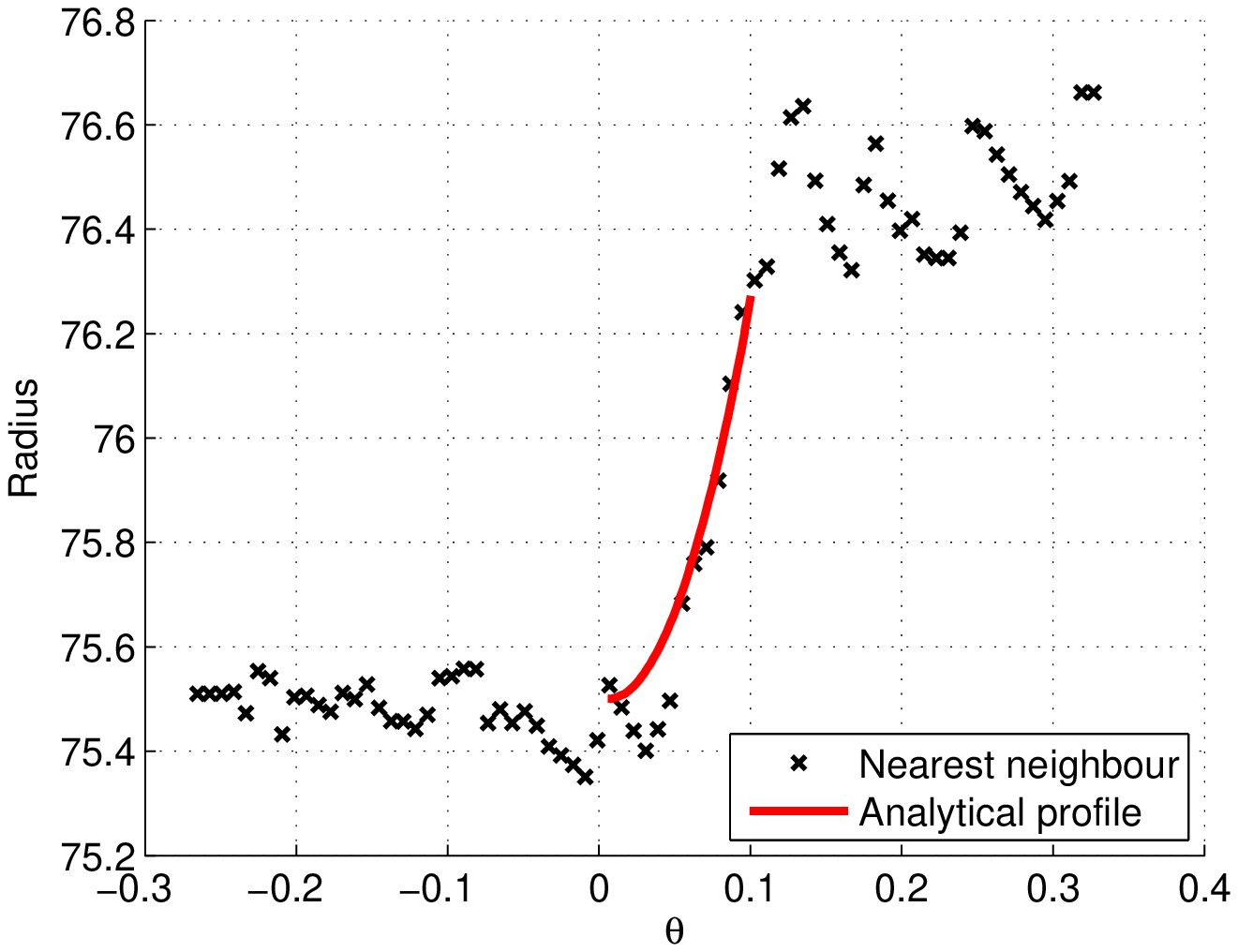}}\\
     %\hspace{.3in}
     \subfigure[]{\label{fig:8b}
          \includegraphics[width=2.35 in]{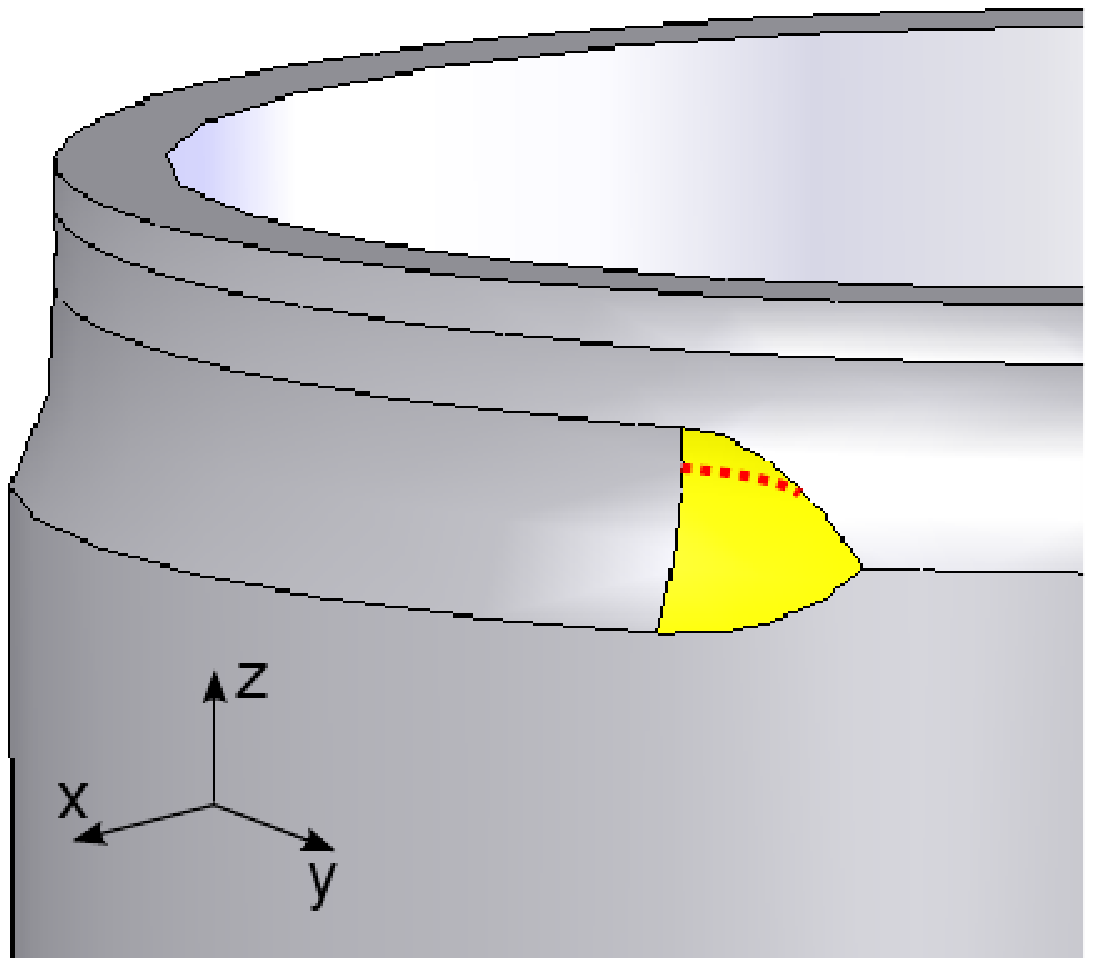}}
\caption[]{Analytical profile of the roller in the best-fit position with the corresponding nearest neighbour experimental points on the (a) $xy$ plane and (b) the location of these points on the relevant portion of the surface in 3D.}
     \label{fig:8}
\end{figure}
\begin{figure}%[htp]
     \centering
     \subfigure[]{\label{fig:9a}
          \includegraphics[width=4 in]{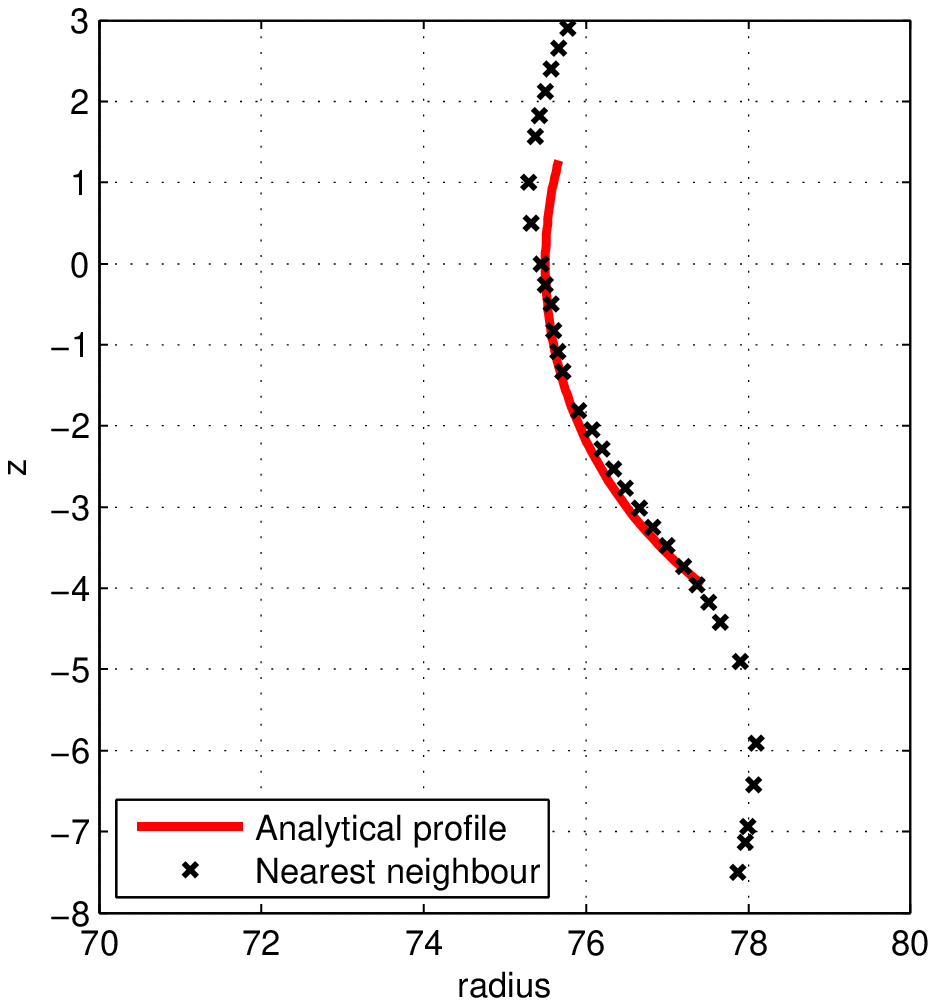}}\\
     %\hspace{.3in}
     \subfigure[]{\label{fig:9b}
          \includegraphics[width=2.35 in]{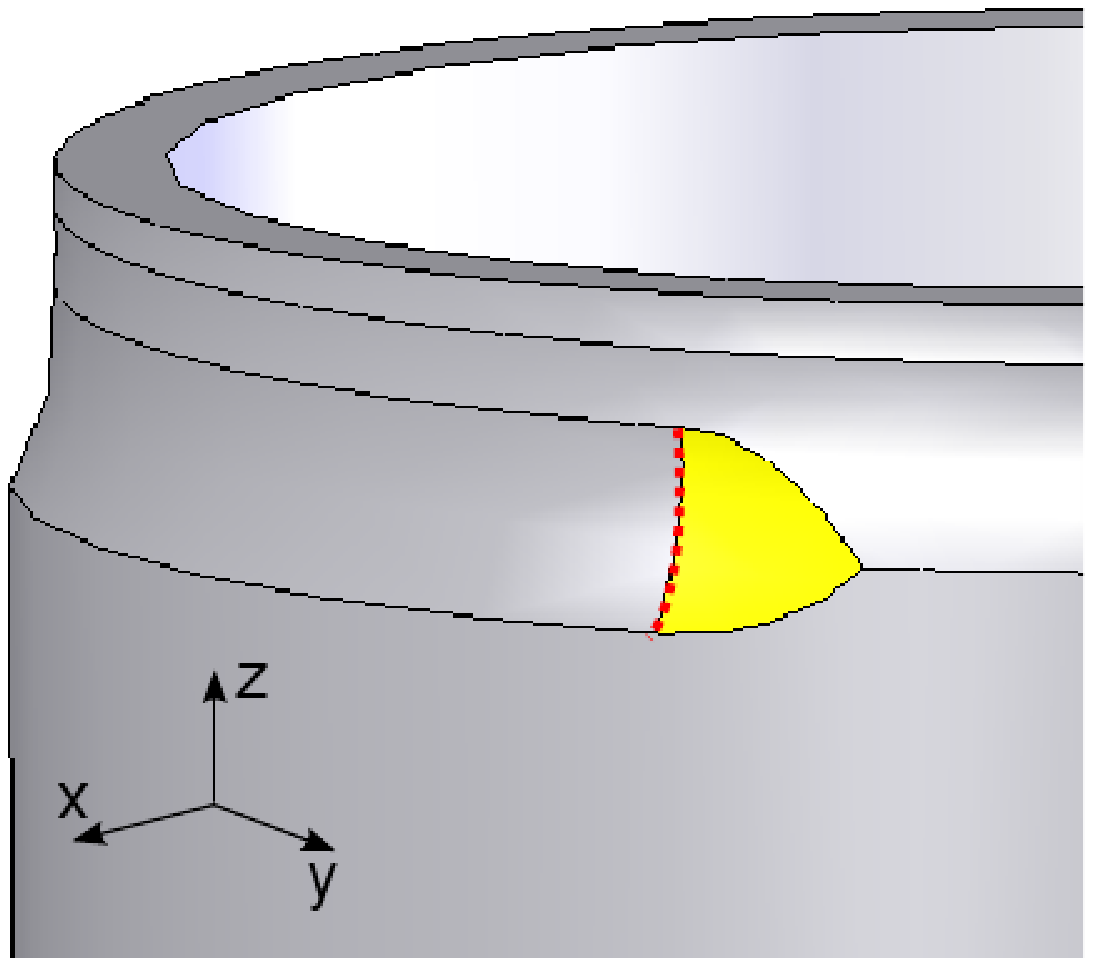}}
\caption[]{Analytical profile of the roller in the best-fit position with the corresponding nearest neighbour experimental points on the (a) $zx$ plane and (b) the location of these points on the relevant portion of the surface in 3D.}
     \label{fig:9}
\end{figure}
Figure 10 is a contour plot of the distance between the analytical and experimental surfaces, $\delta_{Di}$.  This plot shows that the peak distance, approximately 0.35 mm, occurs along the axial entry profile due to the minor bulging of material as it encounters the roller. However, there is very little change in the circumferential $\delta_{Di}$ gradient despite the overall surface variation which re-asserts the accuracy of the fit.
\begin{figure}
\centering
\label{fig:10}
      \includegraphics[width=4 in]{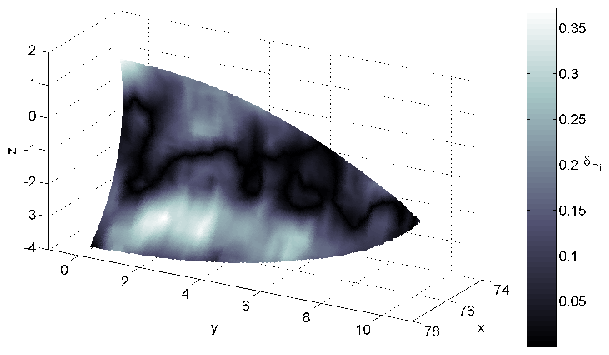}
      \caption[]{Distance between analytical and experimental interpolated surfaces ($\delta_{Di}$) plotted on the analytical surface}
\end{figure}
\section{Application}\label{sec:application}
The flow forming process can be thought of as a simultaneous
combination of both rolling and extrusion (or drawing) processes. An
important application of the calculated roller/workpiece contact
area is the use of the calculated planar projections $A_{xy}$,
$A_{xz}$ and $A_{yz}$ to determine the relative quantities of
extrusion and rolling that are occurring. During flow
forming, there is a set of running conditions that sometimes
generates diametrical growth or defects. This is due to tangential
deformation and, while very small, is approximated with the present analytical solution. \cite{Gur.82} proposed that the ratio of
axial contact length to the circumferential contact length dictates
the ratio of extrusion (or drawing for forward flow forming) to rolling that occurs during flow
forming.  A more accurate measure of this extrusion/rolling ratio
would be to consider the ratios of the $xy$ and the $yz$ projections
of the roller/workpiece contact area.  Therefore, it is important to see how these
quantities vary with respect to the independent process variables.

Using the baseline independent geometric variables presented by \cite{xu.01}, a sensitivity analysis was conducted on the
independent variables that define the three components of the
surface area.  The process values
used by \cite{xu.01} were
$P=0.6$ mm/rev, $t_0=5$ mm, $t_f=3.5$ mm, $R_m=35$ mm,
$R_r=82$ mm, $R=3$ mm and an attack angle of $\alpha
=25^{\circ}$. Note that this set of variables dictates an `AI'
(Section \ref{sec:axiallimits}) contact condition where the exit
angle $\beta$ does not participate. Through implementing One Factor
At a Time (OFAT) analysis, whereby one variable is changed while
holding others constant, the independent variables were changed
individually between -50 and +100\% of the initial values with the exception of $t_0$ and $t_f$.  The starting thickness was varied -25 and +100\% as outside the lower range there is no contact. The final thickness was varied 0 to -50\%. The resultant effects on the
contact area components were then calculated.  These results are presented in Figure 10. The overall effect on changing the independent process variables on both the area and the associated components is given in Table \ref{table:changearea}.
\begin{figure}
\centering
\label{fig:11}
      \includegraphics[width=5.5 in]{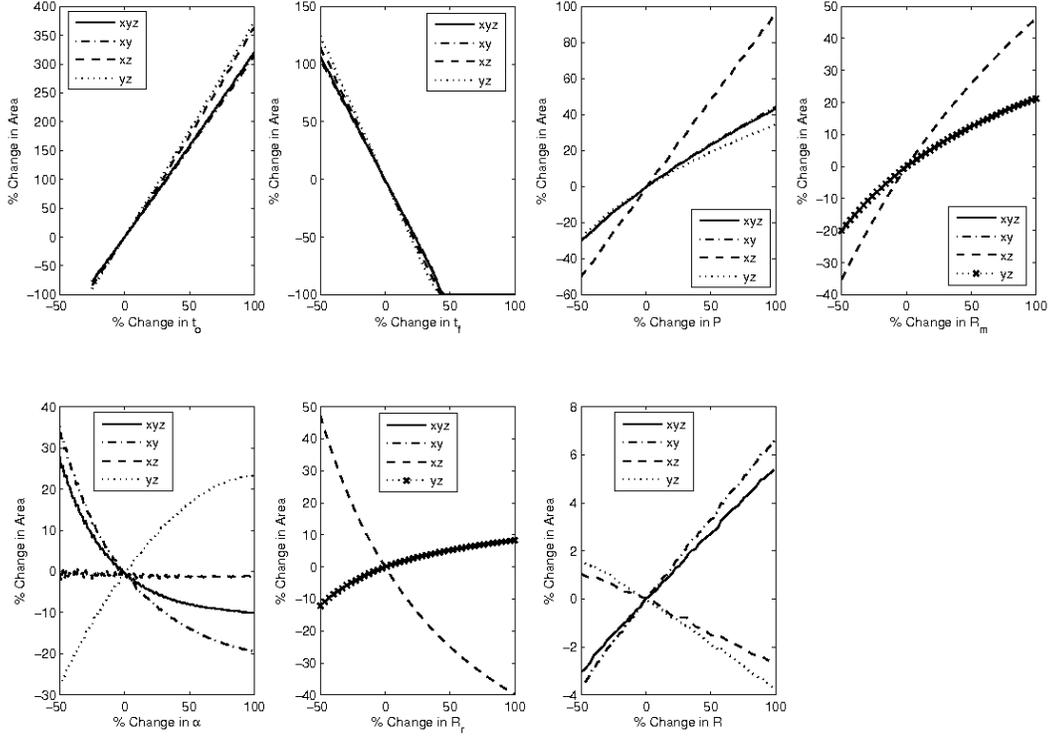}
      \caption[]{The percent change in the contact area
  and projections on major planes by varying a baseline starting thickness ($t_o$), final thickness ($t_f$), pitch ($P$), mandrel radius ($R_m$), attack angle ($\alpha$), roller radius ($R_r$) and roller nose radius($R$). The results for $R_r$ and $R_m$ have $A_{xyz}$, $A_{xy}$ and $A_{yz}$ overlaid.}
\end{figure}
\begin{table}
\centering
\caption{Range of percentage change in contact area with percent change in independent process variables}\label{table:changearea}
    \begin{singlespacing}
    \begin{tabular}{cccccc}
    \hline
    \multirow{2}{*}{\textbf{Variable}} & \textbf{\% Range}  & \multicolumn{4}{c}{\textbf{\% Range in Area}}\\
    & \textbf{in Variable}&\textbf{$A_{xyz}$ } & \textbf{$A_{xy}$} &  \textbf{$A_{xz}$ } &  \textbf{$A_{yz}$}\\
    \hline
%$t_o$ & -77.79 to 319.63 & -76.50 to 313.33 & -83.93 to 363.54 & -89.24 to 372.90\\
%$t_f$ & -100.00 to 106.30 & -100.00 to 104.18 & -100.00 to 114.28 &-100.00 to  124.23\\
%$P$ & -30.10 to  43.25 & -30.33 to  44.15 & -50.01 to  96.30 & -27.80 to  34.44 \\
%$R_m$ & -19.94 to  21.29 & -19.93 to  21.28 & -35.25 to  46.27 & -20.04 to  21.30\\
%$\alpha$ &-10.07 to  27.89 & -19.36 to  35.29 &  -2.53 to   0.67 & -27.57  to 23.42\\
%$R_r$ &  -12.10 to   8.36 & -12.12 to   8.38 & -39.82 to  47.26 & -12.18 to   8.30\\
%$R$  &-3.05  to  5.55 &  -3.65  to  6.74 &  -2.81 to   1.04 &  -3.72 to   1.67 \\
 $t_o$ &125& 397.43 & 389.83 & 447.48 & 462.14\\
$t_f$ &150&  206.30 & 204.18 & 214.27 & 224.23\\
$P$ &150&  73.35  & 74.48 & 146.31 &  62.24\\
$R_m$ &150 &  41.23  & 41.21 &  81.52 &  41.34\\
$\alpha$ &150&  37.95  & 54.65  &  3.20 &  50.99\\
$R_r$ &150& 20.45  & 20.50 &  87.07 &  20.48\\
$R$ &150&  8.60  & 10.39  &  3.85  &  5.39\\
\hline
\end{tabular}
\end{singlespacing}
\end{table}

The OFAT analysis technique is limited as, by definition, it does
not allow for simultaneous changes in multiple variables. For the
given geometry, however, this analysis does display the following
important observations:
\begin{itemize}
\item In terms of the largest effect on the overall contact area $A_{xyz}$, changing the material starting and final thicknesses and the pitch had the largest effect. This is also true for all of the area components, $A_{xy}$, $A_{xz}$ and $A_{yz}$. In order of precedence, the variables that had largest sensitivity on the overall area other than thicknesses and pitch were the radius of the mandrel, the attack angle, radius of the roller, with the nose radius having the least effect overall.

\item Varying the the roller nose radius had the least effect on the contact area as well as the $A_{xy}$ and $A_{yz}$ components.

\item The rolling component, $A_{yz}$, followed the same trends as the overall area for changes in thicknesses, pitch, mandrel/roller radii and attack angle. This component decreased while the extrusion/drawing component and the overall area increased for larger roller nose radii. Furthermore, $A_{yz}$, is more sensitive to the attack angle than the mandrel radius.

\item The largest effect on the $A_{xy}$ component, or the drawing/extrusion part of the deformation saw the same precedence of variables as for the overall contact area. This component showed the same response to variable changes as the overall area.

\item The tangential deformation component, $A_{xz}$, is marginally more sensitive to the radius of the roller than the radius of the mandrel, and the roller nose radius has approximately the same sensitivity as the attack angle. This component also increased for larger values of pitch and mandrel radius, but decreased for larger roller radii and nose radii. $A_{xz}$ remained unaffected by changes in attack angle.

\item The overall contact area increased with increased variable values in all cases except for the final thickness value and the attack angle. This decrease was a linear for the former and non-linear for the latter.

\item The effect of changing the starting thickness, final thickness and roller nose radius is a linear change for all area components while all others are non-linear.
\end{itemize}

These findings are of practical importance to flow forming. If a worn roller is to be re-used after resurfacing, it may be necessary to modify the pitch in order to maintain the same forming geometry when the process was first commissioned. If a single set of rollers are to be used with different mandrels, it is also important from a process design standpoint so that the same forming zone geometry can be maintained. Furthermore, knowing the sensitivity of each of the variables on the overall contact and therefore deformation mode also permits easier troubleshooting of existing processes.

\section{Conclusion}
An analytical model of the roller/workpiece interface in flow forming has been developed such that it may predict the contact area. This model is applicable to all tooling geometries for both forward and backward flow forming processes. Due to the general nature of the description of the geometry, the approach taken can be used for other rotary forming operations where a die or a roller is used to deform a cylindrical workpiece locally. This model has been compared to experimental data generated from physical modelling and shows excellent correspondence. Specifically, the analytical model was found to describe the experimental surface within 0.4 mm based on mean square error.

An example of the application of the model has been demonstrated in the form of a OFAT sensitivity analysis applied to independent geometric variables determining tooling interaction. The independent geometric variables examined were starting and final thicknesses ($t_0$, $t_f$), forming pitch ($P$), mandrel radius ($R_m$), attack angle ($\alpha$) as well as roller and roller nose radii ($R_r$, $R$). These variables were modified over a range of $125\%$ for the starting thickness and 150\% for all others. Specific findings showed that on the basis of a unit change in the respective variables:
\begin{itemize}
\item $t_0$ had four times the effect on the the change in overall area and components
\item $t_f$ had 33\% more of an effect
\item $P$ had 50\% less of an effect, with the exception of the tangential deformation component, $A_{xz}$ which had 50\% more of an effect
\item $R_m$, $\alpha$, and $R_r$ have less than a $27\%$ effect
\item $R$ caused the least change: less than 7\% change in area
\end{itemize}

The present work could be extended to study the multi-variant effects on the contact area to fully account for the geometric changes during complicated forming processes. However, geometric factors are not the only process parameters which govern the process. The main direction of future work is to link the geometric factors to other process factors such as workpiece material properties and tribological considerations to gain deeper insight into the overall process mechanics.

\end{document}